\documentclass[prl,twocolumn,amsmath,amssymb]{revtex4}
\usepackage{graphicx}
\begin{document}
\title{Shear banding phenomena in a Laponite suspension}
\author{F. Ianni$^{1,2}$}
\email{francesca.ianni@phys.uniroma1.it}
\author{R. Di Leonardo$^{2}$}
\author{S. Gentilini$^1$}
\author{G. Ruocco$^{1,2}$}
\affiliation{ $^{1}$ Dipartimento di Fisica, Universit\'a di Roma
``La
Sapienza'', I-00185, Roma, Italy.\\
$^{2}$ SOFT-INFM-CNR c/o Universit\'a di Roma ``La Sapienza'',
I-00185, Roma, Italy. }
\date{\today}

\pacs{83.80.Hj,83.85.Ei,42.25.Fx}
\begin{abstract}
Shear localization in an aqueous clay suspension of Laponite is
investigated through dynamic light scattering, which provides
access both to the dynamics of the system (homodyne mode) and to
the local velocity profile (heterodyne mode). When the shear bands
form, a relaxation of the dynamics typical of a gel phase is
observed in the unsheared band soon after flow stop, suggesting
that an arrested dynamics is present during the shear localization
regime. Periodic oscillations of the flow behavior, typical of a
stick-slip phenomenon, are also observed when shear localization
occurs. Both results are discussed in the light of various
theoretical models for soft glassy materials.
\end{abstract}
\maketitle

\subsection{Introduction}
Under the influence of a shear flow, many complex fluids exhibit
the formation of macroscopic bands, parallel to the flow direction
and characterized by different local shear rates. This phenomenon
is called \textit{shear banding} and is attributed to the presence
of a decreasing branch in the stress-shear rate curve: in a
controlled shear experiment, it is observed when the stress falls
in the interval where the so called flow curve is multivalued.
Experimentally, shear banding has mainly been observed in a class
of complex fluids, like wormlike micelles \cite{berret, britton,
fischer, salmon1}, colloidal crystals \cite{chen}, or lamellar
surfactant systems \cite{wun, salmon2}, where a phase transition
associated to the microstructure occurs under flow. As described
by various theoretical models \cite{spenley, olmsted, lu, dhont,
butler}, in such systems an ordered phase coexists with a
disordered one during the shear banding regime.

However, shear banding has also been observed in soft-glassy
materials \cite{sollich}, like foams \cite{debregeas, laur},
emulsions \cite{coussot}, or glassy suspensions \cite{pignon,
coussot, holmes}, where a homogeneous structure is kept throughout
the system. To our knowledge, few experimental results are
available in the literature for this class of systems \cite{laur,
coussot, pignon, holmes}, while the microscopic origin of the
phenomenon is poorly understood.  Soft-glassy materials are
characterized by a viscosity that increases many orders of
magnitude as time evolves and the gelation process proceeds
(\textit{aging} behavior), and have rheological properties typical
of soft solids, such as solid-like behavior below a finite yield
stress and shear thinning effect \cite{larson}. At the microscopic
level, the gelation process corresponds to a slowing down of the
structural relaxation with the elapsed time, while shear thinning
results from the system structural dynamics being accelerated by a
shear flow (shear \textit{rejuvenation}).

A shear banding behavior is expected for such systems, due to the
presence of a yield stress, that provides a multivalued region in
the flow curve. When heterogeneous flow emerges, a band with null
local shear rate, flowing as a solid block, may coexist with a
band flowing at a finite local shear rate. This behavior has been
called shear localization and, according to numerical and
theoretical models \cite{varnik, picard}, is associated to a
dynamical transition: an arrested dynamics characterizes the
unsheared band, while the sheared band exhibits a liquid-like
behavior. Moreover, these works have evidenced the emergence of a
fluctuating shear banded flow, a phenomenon also observed in non
glassy shear banding systems \cite{fielding}. In order to describe
this fluctuating behavior, a model accounting for intermittent
plastic and elastic events has been developed \cite{picard2,
tanguy}. At the experimental level, a fluctuating shear
banded flow has been observed in glassy suspensions \cite{pignon,
holmes} and in emulsions \cite{debregeas, laur}, while direct
investigation of the system dynamics when shear localization
occurs is still missing.

Among others, proper candidates for the study of the shear
localization phenomena are suspensions of charged anisotropic
colloidal particles such as clay, that have been widely
investigated both for their important industrial application
\cite{olphen} and as a prototype of glassy systems \cite{houches}.
In particular, shear localization have been observed in such
suspensions through magnetic resonance imaging (MRI)
\cite{coussot, bonnrheo} or visualization techniques
\cite{pignon}.

In this paper, we investigate the shear localization phenomenon in
an aqueous clay suspension of Laponite, a highly thixotropic
liquid which undergoes structural arrest. We are both interested
in the study of the system dynamics when the shear bands form and
in the fluctuating flow phenomenon. The technique that we use is
dynamic light scattering (DLS), which provides access both to the
dynamics of the system after shear cessation (in the homodyne
mode) and to the local velocity profile during the flow (in the
heterodyne mode). In particular, once a shear banding profile has
been detected, we monitor the evolution of the system dynamics
soon after flow cessation in the unsheared band. A relaxation of
the dynamics typical of a gel phase is observed, suggesting that
an arrested dynamics characterize the flat band. For the
investigation of the fluctuating flow instead, the heterodyne mode
provides higher spatial and temporal resolution than MRI or other
techniques, previously used to detect the velocity profile when a
shear banding fluctuating flow occurs in glassy suspensions
\cite{holmes, pignon}. We thus observe periodic oscillations in
the flow, which are typical of a stick-slip behavior: a layer
reversibly fractures and reheals, fluctuating between a frozen
state (slip) and a fluidized state (stick). Our results on the
dynamics of the sheared banding system and on the fluctuating
behavior of the velocity profile will show strong analogies with
the elastoplastic model of Ref. \cite{picard2}, which describes
the flow behavior of a yield stress fluid.

\subsection{Materials and methods}
Aqueous Laponite suspensions have been extensively investigated as
a model system for glassy suspensions \cite{kroon, bonnaging,
barbara0} and the rejuvenating effect of a shear flow on the
system dynamics have also been studied \cite{bonnreju, dileo,
fra}. Laponite particles are disk shaped with a diameter of $25$
nm and $1$ nm thickness and get negatively charged on the faces
when dispersed in a polar solvent. For our experiments, Laponite
powder, provided by Laporte Ltd, is dispersed in ultrapure water
at $3\%$ wt concentration and stirred for about 30 min. The
obtained suspension, which is optically transparent and initially
"liquid", is loaded through a $0.45$ $\mu$m filter into a home
made, disk-disk shear cell for DLS measurements. The cell has a
glass disk, of diameter $d=10$ cm, as the rotating plate, and an
optical window as the static plate. The cell gap is $h=7$ mm and
we fix conventionally the $y$ axis along this direction,
corresponding to the velocity gradient direction under flow.

In the set-up we implemented for DLS measurements \cite{velocim},
an incident laser beam (diode pumped solid-state laser,
$\lambda=532$ nm, $P=150$ mW) impinges on the sample passing
through the optical window. The scattered light pass through the
same window and is collected by a mono-mode optical fiber.
Optionally, it can interfere with a coherent local oscillator
field (heterodyne mode) through a fiber collection apparatus.
Collected light is then detected by a photomultiplier and analyzed
by a home made software correlator \cite{fra}. The scattering
geometry is fixed, with a scattering vector
$q=22\;\mu\textrm{m}^{-1}$. We placed the shear cell in order to
have the scattering volume positioned at a radial distance of 2.1
cm from the rotational axis.

The shift of the cell allows us to select the position of the
scattering volume in the cell gap and the possibility of choosing
a heterodyne correlation scheme enables direct access to the
detailed velocity profile \cite{salmon0}. Due to Doppler effect,
the velocity $v$ of the particles in the scattering volume can
indeed be obtained from the frequency $\omega$ of the collected
oscillating intensity in heterodyne mode: $v=\omega/(q
\cos\hat{\boldsymbol{q}\boldsymbol{v}})$, where $q
\cos\hat{\boldsymbol{q}\boldsymbol{v}}$ is fixed by the scattering
geometry. The frequency $\omega$ is simply obtained as the peak
location in the power spectrum of the intensity fluctuations,
acquired in a time interval $0\div T$. The maximum time resolution
that we can achieve with this measurement, at a fixed position of
the scattering volume in the gap, is $T=10^{-2}$ s, while the
spatial resolution is determined by the scattering volume
dimensions $\sim$ 100 $\mu$m.

The dynamics of the system can be investigated through the
normalized correlation function of the scattered intensity
(homodyne mode) $g^{(2)}(t',t)=\langle I(q,t)I(q,t')
\rangle/\langle I(q)\rangle^2$, which is easily represented in
terms of the particle correlation function. In particular,
homodyne DLS directly probes the intermediate scattering function
of the colloidal particles $F_q(t,t')=\langle
\rho_{-q}(t)\rho_q(t') \rangle/\langle |\rho_{q}|^2\rangle$, which
plays a central role in both theoretical and numerical approaches
to glassy dynamics: $g^{(2)}(t,t')=1+|F_q(t,t')|^2$ \cite{berne}.
However, due to geometrical decorrelation effects during the flow \cite{ackerson},
the intermediate scattering function cannot be detected when the system is under shear.
Therefore, we will follow the system dynamics soon
after shear cessation and try to deduce some information on its
dynamical behavior during the flow.
In classical DLS, the correlation function is calculated as an
average over the time-origin. In aging systems, this is possible
when the experimental acquisition time needed to get a good signal
to noise ratio is longer than the characteristic slow relaxation
time of the system $\tau_s$ and shorter than the time one should
wait before changes in $\tau_s$, due to the aging process, are
significant. This condition doesn't hold when the aging dynamics
is characterized by a very fast evolution, as it happens soon
after flow cessation in a shear banding Laponite sample.
Therefore, we cannot obtain the intensity correlation function by
time averaging and an ensemble average over many rejuvenating
experiments is used instead \cite{fra}: $g^{(2)}(t_w, t)=\langle
I(q,t_w)I(q,t_w+t) \rangle_e/\langle I(q, t_w)\rangle_e^2$, where
$\langle..\rangle_e$ indicates the ensemble average over several
intensity evolutions acquired after cessation of a repeated shear
application. Flow stop is taken as the origin of waiting times
$t_w$. We choose the following protocol: a global
shear rate $\dot{\gamma}_1$ is applied to the system for a time
interval $T_1$; after shear cessation, the intensity fluctuations
are collected for a time interval $T_0$ with a time resolution of
$dt$; then a shear rate of the same value $\dot{\gamma}_1$ is
applied for $T_1$ and the cycle starts again. The intensity
autocorrelation function is then calculated in the time window
$dt\div T_0$, as an ensemble average over all the bunches of
counts.

\begin{figure}[h]
\begin{center}
\includegraphics[width=8.5 cm]{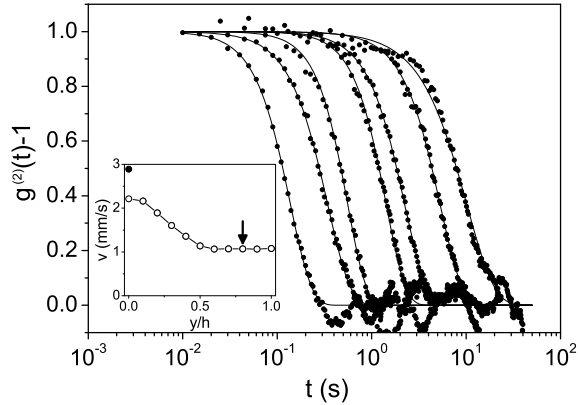}
\caption{Normalized intensity autocorrelation functions measured
in the flat band soon after shear cessation. Correlation functions
are obtained through an ensemble average over many shearing
experiments for 7 logarithmically equidistant waiting times $t_w$
between $3$ s and $67$ s (from left to right) after shear
cessation. A compressed exponential fit
$\exp[-(t/\tau_s)^{\beta}]$ is also plotted for each curve. The
$\beta$ parameter fluctuates between 1 and 2. In the
\textit{inset}, the velocity profile of the fluid during the
experiment is reported and the arrow represents the position of
the scattering volume where the measurements were performed. The
full circle represents the plate velocity to evidence the presence
of wall slip.}\label{dyn}
\end{center}
\end{figure}

\subsection{Results I. Investigating the dynamics}
When a shear rate of the order of 1 s$^{-1}$ is applied to
Laponite samples that have reached dynamical arrest, shear
localization phenomenon is often observed, as shown in the inset
of Fig. \ref{dyn}. Here the velocity profile along the cell gap,
measured through heterodyne DLS, is plotted: a flat band forms
next to the static window, while wall slip occurs at both sides.
According to theoretical models \cite{varnik, picard}, soft glassy
materials exhibiting shear localization are characterized by an
arrested dynamics in the unsheared band. We thus want to
investigate the dynamics of our sample in the flat band.
As we cannot access directly the dynamical
behavior of the system during the flow, we will follow the dynamics in the flat band after shear
cessation and then deduce some information on the shear localization regime.

\begin{figure}[h]
\begin{center}
\includegraphics[width=8.5 cm]{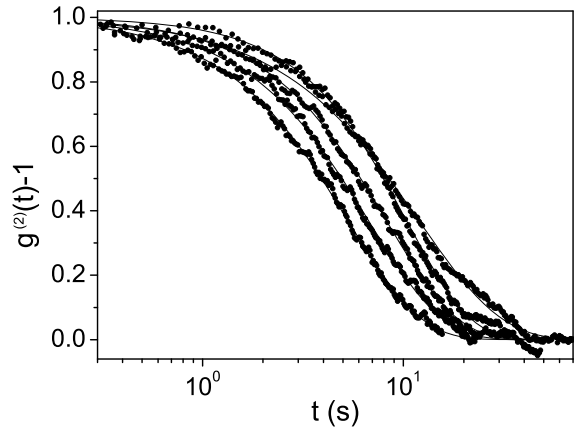}
\caption{Normalized intensity autocorrelation functions measured
in the flat band soon after the application of a small strain
$\Delta x/h=0.3$. Correlation functions are obtained through the
ensemble average procedure for 5 logarithmically equidistant
waiting times $t_w$ between $3$ s and $17$ s after flow cessation.
A compressed exponential fit $\exp[-(t/\tau_s)^{\beta}]$ is also
plotted for each curve. The parameter $\beta$ fluctuates between 1
and 1.5.}\label{strain}
\end{center}
\end{figure}

Through the acquisition method explained above, we monitor the
evolution of the correlation functions with the waiting time since
flow stop, as shown in Fig. \ref{dyn}. These data seem to evidence
an aging behavior after shear cessation, as if the flow had a
rejuvenating effect on the dynamics of the flat band. However, a
null shear rate is not supposed to modify the structure of the
system and induce an acceleration of the structural dynamics
\cite{cates}. In order to exclude that the observed behavior is
due to any residual shear rate in the flat band, we monitor the
dynamics after the application of a small strain to a gelled
sample. In this case, the flow should not modify the structure of
the system, as it is supposed to deform only elastically the
material. Therefore, a strain $\Delta x/h=0.3$ of the duration of
1 s is applied cyclically to a gelled sample and the intermediate
scattering function soon after strain application is measured as
an ensemble average, following the same procedure already
described. As evidenced in Fig. \ref{strain}, an aging dynamics is
still evident after flow cessation, showing that such behavior is
not due to a shear rejuvenating effect. Though the structural
relaxation time scales differently with $t_w$ in the two
experiments, the intermediate scattering functions exhibit a
similar form in the range of $t_w$ here investigated. In
particular, all correlation functions are well fitted by a
compressed exponential: $\exp[-(t/\tau_s)^{\beta}]$ with $\beta>1$
(Fig. \ref{dyn} and Fig. \ref{strain}).

Such behavior of the intermediate scattering function can be
interpreted, in the light of a model developed to describe
anomalous dynamical light scattering in soft glassy gels \cite{bouchaud}, as due to
elastic relaxation of internal stress. In the model, the
system behaves as an elastic medium and force dipoles appear
at random in space and time, inducing micro-collapses of the structure. The decorrelation of light scattered
by the model system is not due to the dynamics of single
scatterers, but to a drift mechanism of big aggregates of
particles and results in a correlation function of the form $\exp[-(t/\tau)^{1.5}]$.
Also an aging behavior is accounted for by
the model, where it is interpreted in terms of strain-dependent
energy barriers. As suggested by this model
for soft glassy gels, the Laponite sample seems to behave as an
elastic medium, which relaxes to equilibrium after the application
of a stress through micro-collapse events. The observed relaxation of the dynamics after
flow cessation is thus coherent with the presence of an arrested (gel) phase
during the flow in the unsheared band.

The investigation of the system dynamics in the sheared band has
already been described elsewhere \cite{fra}. There we monitored,
through the same DLS technique, the evolution of the intermediate
scattering functions after a shear flow is applied to a
dynamically arrested Laponite sample. The same qualitative
behavior is observed in the form of the correlation functions and
in their evolution with $t_w$. This result seems to be in contrast
with the models describing shear localization in soft-glassy
materials \cite{varnik, picard}, where a dynamical transition is
expected: in such models, the unsheared band is characterized by
an arrested dynamics, while the sheared band exhibits a
liquid-like behavior. In the Laponite sample instead, the dynamics
relaxation observed in the sheared band after flow cessation
suggests the presence of a gel phase, as observed in the unsheared
band. The problem may be solved by supposing that, though the
sample is sheared, aggregates of gel phase, larger than the
scattering volume ($\sim$ 100 $\mu$m), exist and slip one over the
other during the flow. After shear, the dynamics in the scattering
volume would thus be characterized by a gel behavior also in the
sheared band. Further experiments should be made to clear up if
such nonhomogeneous flow is effectively present in the sheared
band.

\subsection{Results II. Oscillations in the flow behavior}
\begin{figure}[h!]
\begin{center}
\includegraphics[width=8.5 cm]{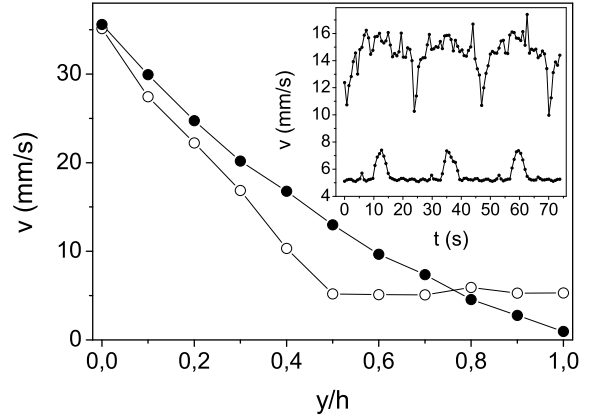}
\caption{A global shear rate $\dot{\Gamma}=4.5$ s$^{-1}$ is
applied to a very old Laponite sample and the velocity profile is
observed to oscillate between the two extreme profiles plotted.
The time period is $T\simeq 23$ s. In the \textit{inset}, the
oscillations of the particle velocity when the scattering volume
is at $y/h=0.4$ (top) and $y/h=0.7$ (bottom) are plotted. The
small and fast oscillations visible in the bottom curve of the
inset have the same period of the disk rotation $T_{rot}\simeq 4$
s. They are due to a slight misalignment of the rotational axis
(with respect to the orthogonal to the window plate), which
induces a small oscillation of the gap width ($\sim
8\%$).}\label{maxminosc2}
\end{center}
\end{figure}

Here we report about the observation of periodic oscillations of
the shear banding velocity profile, exhibited by very old samples.
By monitoring the evolution of the particle velocity through heterodyne DLS, at
different positions along the gap, we are able to rebuild the two
extreme profiles among which the system oscillates. An example is
represented in Fig. \ref{maxminosc2}, where is evident that the
velocity profile oscillates among a configuration where the shear
is localized next to the rotating plate and a configuration with a
linear profile. Most of the time, the system lies in an
intermediate profile between the two. The oscillation period is
larger than the disk rotation period and remains constant on the
timescale of the hours.

Periodic oscillations of the flow behavior when shear localization
occurs have been predicted by the phenomenological model presented in Ref.
\cite{picard} and are here observed for the first time by
directly accessing the velocity profile with a high enough time
resolution. In the theoretical model, such behavior is interpreted as a stick-slip phenomenon:
when shear localization occurs, a diverging viscosity characterizes the unsheared band,
which thus slips on the wall; while in the linear profile configuration,
the system has a liquid-like behavior and sticks on the wall.

\subsection{Conclusions}
Summarizing, the investigation of the shear localization
phenomenon on a Laponite sample have provided two main results:
\textit{i)} through homodyne DLS, the relaxation of the structural
dynamics has been followed in the unsheared band soon after flow
stop and a stress relaxation behavior, typical of a gel phase, has
been evidenced; \textit{ii)} through heterodyne DLS, periodic
oscillations of the shear banding profile, reminiscent of a
stick-slip phenomenon, have been observed. Taking into account
both the oscillating flow behavior and the stress relaxation
mechanism, our experimental results strongly support the
elastoplastic model proposed in Ref. \cite{picard2} to describe
the flow behavior of a yield stress fluid. Two generic ingredients
build up the model, leading to a complex spatiotemporal behavior
of the system: local plastic events occur above a microscopic
yield stress and nonlocal elastic release of the stress follows
\cite{kabla}. At low shear rates, an intermittent flow
localization emerges, with spatially correlated structures forming
parallel to the walls; while at high shear rates the flow is
homogeneous. The stress relaxation behavior and the oscillating
flow localization that characterize this model have both been
observed in our experiments on a Laponite sample. This suggests
that such elastoplastic mechanism takes place in the sample, and
our two main results may be interpreted as two different aspects
emerging from this mechanism.

\end{document}